\documentclass[aps,preprint]{revtex4}%
\usepackage{amsfonts}
\usepackage{amsmath}
\usepackage{amssymb}
\usepackage{graphicx}%
\begin{document}
\title{{\LARGE WHY THE INITIAL INFINITE SINGULARITY OF THE UNIVERSE IS NOT THERE}}
\author{Marcelo Samuel Berman$^{1}$}
\affiliation{$^{1}$Instituto Albert Einstein / Latinamerica - Av. Candido Hartmann, 575 -
\ \# 17}
\affiliation{80730-440 - Curitiba - PR - Brazil email: msberman@institutoalberteinstein.org
\ \ \ \ \ \ \ \ \ \ \ \ \ \ \ \ \ \ \ \ \ \ \ \ \ \ \ \ \ \ \ \ \ \ \ \ \ \ \ \ \ \ \ \ \ \ \ \ \ \ \ \ \ \ \ \ \ \ \ \ \ \ phones:
(+5541)3339-9975 or tele-fax (+5541)3387-1388 \ \ \ \ }
\date{30 January, 2009}

\begin{abstract}
We "explain", using a Classical approach, how the Universe was created out of
\ "nothing" ,i.e., with no input of initial energy. This is a Universe with
no-initial infinite singularity of energy density.



\end{abstract}
\maketitle

\begin{center}
{\LARGE WHY THE INITIAL INFINITE SINGULARITY OF THE UNIVERSE IS NOT THERE}

DOI 10.1007/s10773-009-0007-0

Marcelo Samuel Berman
\end{center}

{\large \bigskip}

{\large \bigskip1 INTRODUCTION}

\bigskip We shall show that in the creation instant of the Universe, the
so-called initial infinite energy density singularity does not exist. In order
to prove that, we take into consideration Robertson-Walker's metric, and find
its associated energy. The reason why the Universe began its existence, while
obeying this metric, and Einstein's field equations, lies in Berman and
Trevisan (2001) hypothesis, that General Relativity theory, though not valid
for making exact calculations inside Planck's Universe, for times less than
\ $10^{-43}$\ sec , due to existence of Quantum uncertainties, reveals the
average\ \ magnitudes of the otherwise uncertain physical values.

\bigskip

\bigskip{\large 2 ZERO-TOTAL ENERGY OF THE UNIVERSE}

\bigskip\noindent Consider Minkowski's metric,

%\bigskip%
$ds^{2}=dt^{2}-\left[  dx^{2}+dy^{2}+dz^{2}\right]  $ \ \ \ \ \ \ \ \ \ \ \ \ \ \ \ \ \ \ \ \ \ \ \ \ \ \ \ \ \ \ \ \ \ \ \ \ \ \ \ \ \ \ \ \ \ \ \ \ \ \ \ \ \ \ (2.1)

%\bigskip

This is an empty Universe, except for test particles. We agree that its total
energy is {zero} (Weinberg, 1972).

%\bigskip

Now consider the expanding flat metric:

%\bigskip%
$ds^{2}=dt^{2}-R^{2}(t)\left[  dx^{2}+dy^{2}+dz^{2}\right]  $\ \ \ \ \ \ \ \ \ \ \ \ \ \ \ \ \ \ \ \ \ \ \ \ \ \ \ \ \ \ \ \ \ \ \ \ \ \ \ \ \ \ \ \ \ \ (2.2)

%\bigskip

Here, $R(t)$ is the scale-factor. At any particular instant of time, $t=t_{0}
$ , we may define new variables, by the reparametrization,

%\bigskip%
$dx^{\prime}{}^{2}\equiv R^{2}\left(  t_{0}\right)  dx^{2}$\ \ \ \ \ \ \ \ \ \ \ \ \ \ \ \ \ \ \ \ \ \ \ \ \ \ \ \ \ \ \ \ \ \ \ \ \ \ \ \ \ \ \ \ \ \ \ \ \ \ \ \ \ \ \ \ \ \ \ \ \ \ \ \ \ \ \ \ \ \ \ \ \ (2.3)

%\bigskip%
$dy^{\prime}{}^{2}\equiv R^{2}\left(  t_{0}\right)  dy^{2}$\ \ \ \ \ \ \ \ \ \ \ \ \ \ \ \ \ \ \ \ \ \ \ \ \ \ \ \ \ \ \ \ \ \ \ \ \ \ \ \ \ \ \ \ \ \ \ \ \ \ \ \ \ \ \ \ \ \ \ \ \ \ \ \ \ \ \ \ \ \ \ \ \ (2.4)

%\bigskip
$dz^{\prime}{}^{2}\equiv R^{2}\left(  t_{0}\right)  dz^{2}$\ \ \ \ \ \ \ \ \ \ \ \ \ \ \ \ \ \ \ \ \ \ \ \ \ \ \ \ \ \ \ \ \ \ \ \ \ \ \ \ \ \ \ \ \ \ \ \ \ \ \ \ \ \ \ \ \ \ \ \ \ \ \ \ \ \ \ \ \ \ \ \ \ \ (2.5)\ \ \ \ \ \ \ \ \ \ \ \ \ \ \ \ \ \ \ \ \ \ \ \ \ \ \ \ \ \ \ \ \ \ \ \ \ \ \ \ \ \ \ \ \ \ \ \ \ \ 

%\bigskip

$dt$'$^{2}\equiv dt^{2}$

\begin{center}

\end{center}

%\bigskip

Then,

%\bigskip

$ds^{2}=dt^{\prime2}-\left[  dx^{\prime2}+dy^{\prime2}+dz^{\prime2}\right]  $\ \ \ \ \ \ \ \ \ \ \ \ \ \ \ \ \ \ \ \ \ \ \ \ \ \ \ \ \ \ \ \ \ \ \ \ \ \ \ \ \ \ \ \ \ \ \ \ \ \ \ \ \ \ \ \ \ \ \ \ \ \ \ \ \ \ \ \ \ (2.6)

%\bigskip

The energy of this Universe is the same as Minkowski's one, namely, \ \ $E=0$.
We remember that in energy calculations, the instant of time is fixed.

\bigskip\bigskip
%\bigskip
%\bigskip
%\noindent{\LARGE\bf Conclusion}
%\bigskip
%\bigskip

\noindent\bigskip\noindent Consider now the metric:

%\bigskip%
$ds^{2}=dt^{2}-\frac{R^{2}(t)}{\left[  1+\frac{kr^{2}}{4}\right]  ^{2}}\left[
dx^{2}+dy^{2}+dz^{2}\right]  $\ \ \ \ \ \ \ \ \ \ \ \ \ \ \ \ \ \ \ \ \ \ \ \ \ \ \ \ \ \ \ \ \ \ \ \ \ \ \ \ \ \ \ \ \ (2.7)

%\bigskip

Here, \ \ $k=0$ \ \ \ yields the flat case, already studied. When \ \ $k$ =
$\pm1$, \ \ \ we have finite closed or infinite open Universes.

%\bigskip

We want to calculate its energy. We are allowed to choose the way into making
the calculation, so we choose a fixed value $\bar{r}$ of the radial
coordinate, for which we reparametrize the metric:\bigskip

\bigskip
%\bigskip%
$dx^{\prime i}{}^{2}\equiv\frac{R^{2}\left(  t_{0}\right)  dx^{i2}}{\left[
1+k\frac{\bar{r}^{2}}{4}\right]  ^{2}}$\ \ \ \ \ \ \ \ \ \ \ \ \ \ \ \ \ \ \ \ \ \ \ \ \ \ \ \ \ \ \ \ \ \ \ \ \ \ \ \ \ \ \ \ \ \ \ \ \ \ \ \ \ \ \ \ \ \ \ \ \ \ \ \ \ \ \ \ \ \ \ \ \ (2.8)

\ 

\ \ \ \ \ \ \ \ \ \ \ \ \ \ \ \ \ \ \ \ \ \ \ \ \ ( $i=1,2,3$\ )

\bigskip

For this value of \ \ $r=\bar{r}$, the reparametrized metric has zero energy
value, by the same token as above. Now we sum for all other values of \ $r$\ ,
obtaining an infinite sum of zeros, which yields a total energy of zero value.\ \ 

\bigskip
%\bigskip
%\bigskip
%\noindent{\LARGE\bf Conclusion}
%\bigskip
%\bigskip

\noindent

{\large 3 SINGULARITY-FREE UNIVERSE}

\bigskip

\bigskip Now, we can check the creation instant, where \ $t\rightarrow0$\ .
Consider that, from Einstein's field equations, we have found a scale-factor
that obeys the condition,\bigskip

$\qquad\qquad\qquad\qquad\qquad\qquad\lim\limits_{t\rightarrow0}$ $R(t)=0$\ \ \ \ \ \ \ \ \ .\ \ \ \ \ \ \ \ \ \ \ \ \ \ \ \ \ \ \ \ \ \ \ \ \ \ \ \ \ \ \ \ \ \ \ \ \ \ \ \ \ \ \ \ \ \ \ \ \ (3.1)

\bigskip

\bigskip On the other hand, we are going to calculate the energy density,
which is usually defined as,

\bigskip

\ \ \ \ \ \ \ \ \ \ \ \ \ \ \ \ \ \ \ \ \ \ \ \ \ \ \ \ \ \ \ \ \ \ \ \ \ \ $\rho
=\frac{E}{V}$ \ \ \ \ \ \ \ \ \ \ \ \ \ \ . \ \ \ \ \ \ \ \ \ \ \ \ \ \ \ \ \ \ \ \ \ \ \ \ \ \ \ \ \ \ \ \ \ \ \ \ \ \ \ \ \ \ \ \ \ \ \ \ \ \ (3.2)

\bigskip From energy conservation, the fact that \ \ $E=0$\ \ is an
independent of time result. In the above equation, \ \ $V$\ \ \ stands for
tri-dimensional volume, \ \ \ \ 

\bigskip

\ \ \ \ \ \ \ \ \ \ \ \ \ \ \ \ \ \ \ \ \ \ \ \ \ \ \ \ \ \ \ \ \ \ \ \ \ \ $V=\frac
{4}{3}\pi R^{3}$\ \ \ \ \ \ \ \ \ \ \ \ \ \ \ \ \ \ \ . \ \ \ \ \ \ \ \ \ \ \ \ \ \ \ \ \ \ \ \ \ \ \ \ \ \ \ \ \ \ \ \ \ \ \ \ \ \ \ \ \ (3.3)

\bigskip

The so-called initial infinite singularity is in fact, from what we have
calculated above, a kind of indeterminated relation of the type \ \ $\frac
{0}{0}$\ \ \ , when \ \ \ \ $t=0$\ \ \ \ . However, there is a theorem about
limits, that says that when the limit to the left is equal to the limit to the
right, the limit is equal to that result. In other words,

\bigskip

$\qquad\qquad\qquad\qquad\lim\limits_{t\rightarrow0^{-}}$ $\rho(t)=\lim
\limits_{t\rightarrow0^{+}}$ $\rho(t)=\lim\limits_{t\rightarrow0}$ $\rho
(t)=0$\ \ \ \ \ \ \ \ \ \ \ \ .
\ \ \ \ \ \ \ \ \ \ \ \ \ \ \ \ \ \ \ \ \ \ \ (3.4) \ \ \ \ \ \ \ \ \ \ \ \ \ \ \ \ \ \ \ 

\bigskip

[ The first two left-hand side terms in the multiple equality, are zero
because they result from the fraction $\ \ \ \frac{O}{V}=0$\ \ \ , because
\ \ \ \ $V^{-}$ $\ \ $\ and$\ \ \ \ \ V^{+}$ $\ \ \ \ $are$\ \ \neq0$\ \ \ \ ].

\bigskip

\bigskip The reason for a zero-total energy density, lies in the fact that the
positive energies, which are constituted by the sum of visible and dark
matter, radiation, cosmological "constant" (dark energy) and other fields,
must be added by an equal amount of (negative) potential energy density.

It is now evident that we have no infinite singularity at the origin of time.

\bigskip

{\large 4 CONCLUSIONS}

\bigskip

The above original result, seems not to have been considered by any author. We
obtained, from Robertson-Walker's metric, a scale-factor that begins from
zero-value at the initial time, while we showed that the initial energy and
energy density, are equally zero.

\bigskip

The basis of our calculation has been the zero-total energy of the
Robertson-Walker's Universe. It can be checked that pseudo-tensor calculations
yield the same result for the total energy (Berman, 2006; 2006a; 2008). Other
author calculations of the energy run similarly (Cooperstock and Israelit,
1995; Feynman, 1962-63). Inflationary cosmology also recovers a zero-total
energy of the Universe (Guth, 1981).

\bigskip

\bigskip\bigskip{\Large Acknowledgements}

\bigskip I thank my intellectual mentor Prof. Fernando de Mello Gomide, and,
Nelson Suga, Marcelo F. Guimar\~{a}es, Antonio F. da F. Teixeira, Mauro
Tonasse, and I am also grateful for the encouragement by Albert, Paula, and
Geni. I offer this paper \textit{in memoriam of \ \ }M. M. Som

\bigskip

{\Large References}

\bigskip

Berman, M.S (2006) -- \textit{Energy of Black-Holes and Hawking's Universe.
}In \textit{Trends in Black-Hole Research, }Chapter 5\textit{.} Edited by Paul
Kreitler, Nova Science, New York.

\begin{description}
\item Berman, M.S. (2006a) - \textit{Energy, Brief History of Black-Holes, and
Hawking's Universe. }In \textit{New Developments in Black-Hole Research},
Chapter 5\textit{.} Edited by Paul Kreitler, Nova Science, New York.
\end{description}

\bigskip Berman,M.S. (2008) - \textit{A Primer in Black-Holes, Mach's
Principle, and Gravitational Energy }, Nova Science, New York.

Berman, M.S.; Trevisan, L.A. (2001) - \ \textit{On the Creation of the
Universe out of "nothing"}\ , Los Alamos Archives http://arxiv.org/abs/gr-qc/0104060

Cooperstock, F.I; Israelit, M. (1995) -- Found. of Physics, 25:(4), 631.

Feynman, R.P.(1962-63) -- \textit{Lectures on Gravitation}, \ class notes
taken by F.B.Morinigo; W.C. Wagner, Addison Wesley, Reading.

Guth,A. (1981) --Phys Rev D \textbf{23}, 247.

Weinberg, S. (1972) - \textit{Gravitation and Cosmology, }Wiley, New York.

\end{document}